\documentclass{sigchi-ext}
\usepackage[T1]{fontenc}
\usepackage{textcomp}
\usepackage[scaled=.92]{helvet} 
\usepackage{graphicx} 
\usepackage{balance}  
\usepackage{booktabs} 
\usepackage{ccicons}  
\usepackage{ragged2e} 
\usepackage{tikz}
\usepackage{color}

\def\plaintitle{Liveness in Interactive Systems} 
\def\emptyauthor{}
\def\plainkeywords{Authors' choice; of terms; separated; by
  semicolons; include commas, within terms only; required.}

\title{Liveness in Interactive Systems}

\numberofauthors{1}
\author{%
    \textbf{Sang Won Lee}\\
    \affaddr{Virginia Tech} \\
    \affaddr{Blacksburg, VA 24060, USA} \\
    \email{sangwonlee@vt.edu} } 

\hypersetup{%
  pdftitle={\plaintitle},
  pdfauthor={\emptyauthor},
  pdfkeywords={\plainkeywords},
  bookmarksnumbered,
  pdfstartview={FitH},
  colorlinks,
  citecolor=black,
  filecolor=black,
  urlcolor=linkColor,
  breaklinks=true,
}

\begin{document}
\maketitle

\CopyrightYear{2018}
\doi{https://doi.org/10.5281/zenodo.1471026}
\copyrightinfo{\acmcopyright}
\RaggedRight{} 

\begin{abstract}
Creating an artifact in front of public offers an opportunity to involve spectators in the creation process. 
For example, in a live music concert, audience members can clap, stomp and sing with the musicians to be part of the music piece. 
Live creation can facilitate collaboration with the spectators. 
The questions I set out to answer are what does it mean to have liveness in interactive systems to support large-scale hybrid events that involve audience participation. 
The notion of liveness is subtle in human-computer interaction.
In this paper, I revisit the notion of liveness and provide definitions of both \textit{live} and \textit{liveness} from the perspective of designing interactive systems.\footnote{The large portion of this paper appear in \cite{lee2018thesis}.} In addition, I discuss why liveness matters in facilitating hybrid events and suggest future research works. 
\end{abstract}


\category{See}{\url{http://acm.org/about/class/1998/}}{for
  full list of ACM classifiers. This section is required.}

\section{Introduction}
Creating an artifact --- such as writing a book, developing software, or performing a piece of music --- is often limited to those with domain expertise. 
As a consequence, effectively involving non-expert end users in such creative processes is challenging.
One potential solution to this is to create an artifact \textit{live} and to involve spectators in the creation process guided by the creators. 
In live creation, creators reveal the process of creation to spectators and the immediate and continuous visibility --- perceptibility, to be more precise --- of the live creation process helps non-experts have better understanding on the process (and the artifact) and be able to participate in the process. 
While increasing number of HCI literature use the term ``live'' interaction and ``liveness'', the notion of liveness is yet subtle in designing and evaluating interactive systems. 
I have explored \textbf\textit{{liveness}} in various interactive systems to involve non-expert users for applications such as programming, writing, music performance, and UI design~\cite{chen2017codeon,  lee2016crowd,lee2015live, lee2018appar, lee2017sketchexpress}.
Through the previous works that I explored, I suggest technical definitions of both ``live'' and ``liveness' that can be used in designing and evaluation interactive systems.
Also I discuss the challenges of involving users in live creation and how technologies can address such challenges. 

\section{Definition of \textit{Live}}

In the Oxford Dictionary, the adjective \textit{live}, in this context, means ``relating to a musical performance given in concert, not on a recording'' or ``transmitted at the time of occurrence, not from a recording''.
In the Merriam-Webster Dictionary (online), \textit{live} is defined as ``of or involving a presentation (such as a play or concert) in which both the performers and an audience are physically present'' or ``broadcast directly at the time of production''. 
Two common components of the above definitions are the concurrence of action and perception, and the notion of an artifact that is produced, performed, or transmitted, such as a musical performance, TV show, or sporting event. 
The concurrence of action and perception implies live settings include two different groups: creators --- those who are in action (performers, actors, broadcasters) --- and users (consumers more broadly) --- those who perceive the action (an audience, viewers, listeners). 
In summary, there exist four components that constitutes being live: 1) an artifact, 2) those who create/deliver the artifact, 3) those who perceive it, and 4) the concurrence between 2 and 3.

I define the term \textit{\textbf{live}}  as follows: 
\\
\vspace{0.5cm}
\fbox{%
    \parbox{0.45\textwidth}{%
\textit{relating to the process of creating or delivering an artifact being \textbf{perceptible in (near) real time} to spectators}. 
    }%
}
\\
\vspace{0.5cm}

First, the term ``perceptible'' is used in its broadest sense to include any capability of being perceived through any human sensory system (e.g., visible, audible, tangible, smellable, tastable).
For example, a live concert can be heard over the radio. 
Frequently, the term live relates to the perception of multisensory information, typically audiovisual but can extend to other senses (taste, smell, touch). 
The process being perceptible can be naturally accomplished when creators and spectators are co-located at the same time. 
In this case of live creation, the co-locatedness includes the spectators have access to  --- can perceive --- the process of creation.
However, it does not necessarily mean that live creation needs to happen only in co-located setup. 
Creators and spectators can be remote and the tele-communication technology that is used to make the process perceptible will be required. 
Interestingly, the notion of liveness has developed with the advancement of media technologies through which one can transmit audiovisual information to a remote location (radio, ``tele''phone, ``tele''vision, and recording) 
For example, before recording technology, there was no way to listen to music other than live performance; all music was live music back then. 

Second, being perceptible in ``(near) real time'' means that the process is perceptible at the time of occurrence or with minimal delay. 
There is no transmission delay involved for the live process for which spectators and performers are co-located is physically perceptible in real-time.
Therefore, the term ``(near) real time'' typically relates to the latency involved in technologies that enable the perceptibility of the process and artifacts to remote spectators.
The notion of (near) real time is defined as the timeliness of data in the context of distributed systems and can depend on the situation, implying that there are no significant delays~\cite{federal1996federal}. 
For example, having a long latency is even desirable in certain context; in typical live TV shows, a broadcast delay is intentionally added to allow undesirable content, such as profanity or nudity, to be censored~\cite{conrad2008fleeting}.
On the other hand,in the case of an interactive system (two-way communication between creators and spectators) beyond media delivery, latency can negatively impact the user experience. 
For example, a teleconferencing system (such as Skype) allows creators to present the process of creation live to users, but a delay of more than a few seconds would make two-way communication difficult~\cite{ralston2000encyclopedia}.

\section{\textit{Live} Creation}

Typically creation processes are not live --- in fact, they are often \textit{hidden} to users (or consumers more broadly) entirely.
The creation of an artifact takes place asynchronously with respect to its consumption, and the process of creation is separated from the users. 
For example, readers (users) read a book (an artifact) only after the author (its creator) has finished writing it. 
The same is typically true of food that people eat at a restaurant, a painting exhibited at a museum, or software written by developers.
On the contrary, in live settings, the process of creating artifacts is revealed to users to an extent that is typically understandable and presentable to spectators. 
For example, at a live music concert, music is the \textit{artifact}, musicians are the \textit{creators}, and audience members are the \textit{users}. 
While many aspects of a music performance are still asynchronously done or hidden --- such as practices, rehearsals, back-stage efforts --- the audience get to see the actual performance and perceive it as a live creation of an artifact. 
Such asynchronous efforts are more necessary for especially making the process live.

Many creators choose to reveal the process of creation to users so that users can understand how an artifact is made and appreciate the effort that goes into creating an artifact. 
In general, people put more value on an artifact that is created live. 
For instance, people are willing to pay more for live music performances, even though listeners are less and less willing to pay for recorded music~\cite{doi:10.1177/1029864916650719}. 
In the case of cooking, in some restaurants, chefs cook in front of their customers in real time. This is true of teppan-yaki restaurants (sometimes referred to as ``Japanese steakhouses'' in the US), which serve a style of Japanese cuisine cooked on an iron hot plate called a \textit{teppan}~\cite{wiki:Teppanyaki}. 
The visible process of creation adds value to the artifact because users can understand how the artifact is created~\cite{auslander2008liveness}. 

Live creation encourages creators to reinforce the performative aspect of the creation process.
Creators would want to augment their actions (i.e., narration) and to add even unnecessary steps or modifications (i.e., exaggerated movements of teppan-yaki chef) for make the process more transparent and engaging. 
The perceptibility on the artifact to spectators and the existence of them allows creators to express their thoughts and emotion on the artifact and to convey their creative practice. 
Increasing such demonstrative components of a live interaction gives creators opportunities to engage people through understanding and theatrical elements. 
For users standpoint, as perceiving the process takes time, the artifact now is something that can be not only used but also experienced. 
Creators would ``perform'' the creation activity live like performing arts compared to the case of non-live creation with no audience. 

\section{Definition of \textit{Liveness}}\label{sec:defliveness}

How, then, can we define \textit{\textbf{liveness}}? According to the Oxford Dictionary  liveness is ``the quality or condition (of an event, performance, etc.) of being heard, watched, or broadcast at the time of occurrence.''
In general, liveness has been valued, especially in the context of performing arts such as music, theatre, and dance.  
Auslander explored the value of liveness, particularly in the era of culture dominated by (typically non-live) mass media and media technology, such as television~\cite{auslander2008liveness}. 
Liveness also has been expanded to broader areas digital arts and new media. 
Crisell also states that a broadcast conveying messages over distances without the time lapse is the basis of the broadcast's liveness~\cite{crisell2012liveness}. 

A non-live process can still incorporate liveness through the use of techniques that give the audience \textit{the sense of being there} and the opportunity to witness what happens. 
A typical example of a non-live creative process that still has liveness is replay of recorded live events. 
Based on the definition of live used in the previous section, recording of a live music performance is NOT live due to the delay between creation and consumption.  
However, it was live at the time of recording, and the recording effectively shows the characteristics of live settings: a number of characteristics unique to live settings can be found, such as any risks involved, the impromptu nature of the performance, the sound of the audience cheering, and the unedited visuals, especially compared to non-live forms of them --- studio-recorded music. 
Hook et al. describe liveness as ``the properties of intimacy and immediacy experienced by both spectators and performers'', which comes from the spectators' proximity to and even inclusion in a performance of some kind~\cite{Hook:2012:EHR:2212776.2212717}. 
The authors assert that the uses of technology can bring a sense of presence and involvement in events, which can exist independently of time and space. 
For example, having multiple views of a live music concert on a television is impossible to replicate for an audience member at a concert hall. 
Therefore, liveness can be related to the qualities and properties that help users experience the process of creating an artifact as if they are there.

I define the term \textit{\textbf{liveness}}  as follows: 
\\
\vspace{0.5cm}
\fbox{%
    \parbox{0.45\textwidth}{%
\textit{the extent to which the process of creating artifacts and the state of the artifacts are \textbf{immediately and continuously perceptible}}%
    }%
}
\\
\vspace{0.5cm}

\noindent This particular definition is selected in consideration of the context of interactive systems design and can be used to quantitatively assess liveness of an interactive system.
Liveness can be considered in terms of three values: \textit{immediacy}, \textit{continuity}, and \textit{perceptibility}. 
In the following subsections, I will discuss each quality of liveness.

\subsection{\textbf{Immediacy: minimizing the latency}}
\begin{figure}
    \centering
    \hspace{-15pc}
    \includegraphics[width=1.5\linewidth]{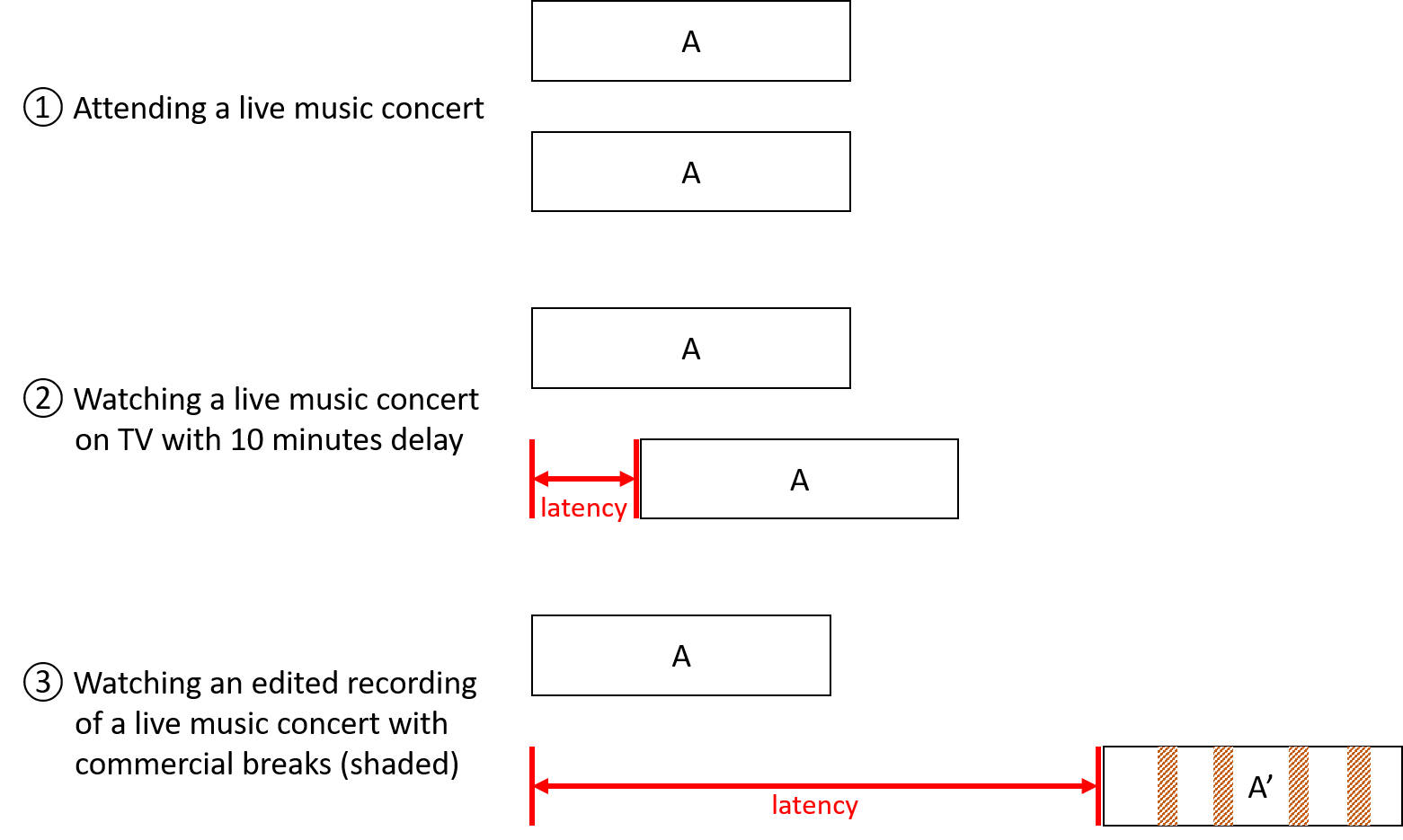}
    \caption{Immediacy and continuity: two temporal dimensions of liveness (1) \textit{live}: an example in which the process of creation is immediately and continuously perceptible, (2) an example in which the process of creation is not immediately but continuously perceptible, the latency can be a measure that can represent the immediacy, (3) an example in which the process of creation is neither immediately nor continuously perceptible. The shaded portion that the state of A and the state of A' differ can be a measure that represents the continuity. }~\label{fig:diagram}
  \end{figure}

The most straightforward quality of liveness is immediacy. 
The immediacy can be simply measured by the average latency between the time of creation and the time that the process is perceptible to spectators. 
For example, live broadcasting of a music concert has more liveness than replaying a recording of the same concert(See Figure-\ref{fig:diagram}-2), as the former becomes perceptible sooner after the time of creation. 

The value added by revealing the process of creation immediately can vary depending on the artifact that is created. 
In general, the artifact of which its value increases with recency --- minimizing the delay between creation and consumption ---  can have potential increase in its value with live creation. 
In some cases, the recency itself can have a value in its quality --- e.g., freshly cooked vs. microwaved food. 
The immediacy requirement can result from a highly personalized nature of artifacts usage, services, or content --- e.g., mass-produced ready-made products(ready-made clothes) vs. made-to-order products(order-made dress). 
In these case, being able to defer the creation until the time to get feedback from the consumers is important values that therefore immediacy brings
The immediacy is an essential quality that enable real-time interaction between spectators and creators as discussed in a previous section. 
Lastly, the value of immediacy can also come from the value of uncertainty involved in the live process --- e.g., live broadcast of sports events vs. recording of them given the final result is known --- and the creators' efforts to reconcile such risks --- e.g., live music performance vs. videotaped performance in a studio which could have done in multiple takes.

\subsection{\textbf{Continuity in revealing the creation process}}

The continuity (or continuous perceptibility) is how continuously the state of the process is perceptible to spectators. 
Continuity can account for why recordings of live events can still have liveness. 
If there is an artifact whose state is \texttt{A}, the state visible to spectators \texttt{A'} can be updated discretely or intermittently. 
The discrepancy between \texttt{A} and \texttt{A'} may arise not only from the latency, but also from the discontinuous visibility (or state synchronization) coming from the mediating technology (See Figure-\ref{fig:diagram}-3). 
Continuity can be orthogonal to immediacy of an interactive system or live events.
For example, a live TV sporting event with commercial breaks is not continuously visible to viewers, because viewers cannot see what happens during the breaks, unlike the audience attending the event in person.
Text editors are another software example. 
Suppose one wants to write text live to remote viewers, for educational purposes in a shared editor (such as Google Docs). 
In this shared editor, the document is continuously shared --- the state is synchronized with every keystroke --- even though there can be some latency. 
However, in the case of an instant chat messenger, messages are shared only when a user presses the Return key (or presses a send button); until that point, the typed text is not visible to the other party. 
In this case, the artifact (conversation) is shared discontinuously. 
Therefore, Google Docs has more liveness than an instant chat messenger. 
The notion of continuity is particularly effective to understand the liveness when certain media or processes are not live and do not occur in real time --- a recording of a live event, for instance. 

As an extreme example, a film can be seen as a medium that has less liveness compared to a play in a theatre.
Based on the definition used here, in terms of immediacy, the final artifact may be presented to viewers with months and years of latency. 
Another aspect that makes film have less liveness is the discontinuous ways in which the artifact is created and presented.
The total time it takes to produce a film may span years. 
However, only a tiny portion of the entire creative process is visible to the audience, as the typical running time of a film is from two to three hours.
However, it takes one hour to perform an hour-long play, and it is visible for an hour to the audience. 
While a tremendous amount of pre-production process may be required (writing, scripting, rehearsal) for a theatrical performance, the fact that the one-hour process is continuously visible to the audience in real time since the beginning constitutes liveness of a theatrical performance and brings with it the impromptu and risky nature of live performance. 
Similar kinds of liveness coming from continuity can exists in asynchronous filming.
For example, a long take in a film can increase the liveness for a particular scene. 

\subsection{\textbf{Perceptibility: the sense of being there}}

Lastly, perceptibility is an important characteristic of liveness. 
It is related to creating \textit{the sense of being there}, typically when it is not, thus the spectators can see, hear, and feel the creative process.
Therefore being co-located in real-time does create a natural perceptibility. 
However, the sense of being there can be augmented with technology. 

In addition, the notion of perceptibility can dynamically change over time with the emergence of new media. 
Virtual reality can be used to create audiovisual perceptibility and many researchers are working on supplementing the other senses as well~\cite{burdea1996force,sutcliffe2003multimedia,Ranasinghe:2017:VVC:3123266.3123440}.  
Revisiting the example of film, a film may bring more liveness by augmenting the sense of being there, not only through well-made content, but also by media technologies, such as a surround screen\cite{Cruz-Neira:1993:SPV:166117.166134}, 3D cinema~\cite{mendiburu20123d}, spatialized audio~\cite{rumsey2012spatial}, and haptic feedback~\cite{danieau2014toward}. 
The technological space of enhancing perceptibility is dynamically expanding and challenging to define. 
These two temporal dimensions --- immediacy and continuity --- provide metrics for us to evaluate and compare interactive systems in terms of liveness.
At the moment, there is no clear way to measure perceptibility other than asking the participants the perceived value, which can vary across individuals. 

One underlying requirement for the perceptibility is that spectators should be able to understand the creation process from their perception. 
While the perceptibility does not include the extent to which the process of creation is easy to understand for the spectators, the effect of liveness will be minimal if the spectators cannot comprehend the process of creation. 
Research in live electronic music, in which performers often sit behinds a laptop computer and an audience is not aware of what is going on, highlights the challenge of delivering liveness that is decoupled from a performer's physical actions~\cite{croft2007theses,emmerson2017living}.

\section{Liveness for Hybrid Events}

Hybrid events do not necessarily need technical components to realize. However, having computational supports can enable novel audience participation modes that were not available before\cite{Cerratto-Pargman:2014:UAP:2639189.2641213, Nelimarkka:2016:LPA:2901790.2901862, Peltonen:2007:ELE:1329469.1329487}. 
For example, I have developed two interactive systems for audience participation in music concerts~\cite{lee2016crowd, lee2013echobo}. 
Typically, audience participation in music concerts was limited to the level in which the audience generated sound only accompanies the primary sound coming from the stage. 
In addition, audience participation is limited to musicians whose music is well known to their audience.
On the contrary, the systems that I developed enable immediate participation and the only sound of a music piece is coming from the audience, not from the on-stage musician(s). 
In these settings, liveness is an important virtue for their participation as having liveness allow the participants to eventually understand how their participation contributes to creating an artifact.
While the co-located nature of such event naturally brings liveness, using interactive systems to facilitate co-located collaboration sometimes add ruin to the liveness by hiding the process of creation with information lost in digitized participation. 
However, having liveness can be costly as it requires significant engineering efforts for multi-user software: synchronizing the state in the finest resolution for the continuous perceptibly, minimizing the latency, and being able to perceive creators actions entirely.
Therefore, I suggest two types of research in understanding the effects of liveness in facilitating hybrid events. 

\begin{marginfigure}
  \begin{minipage}{\marginparwidth}
     \centering
    \includegraphics[width=\linewidth]{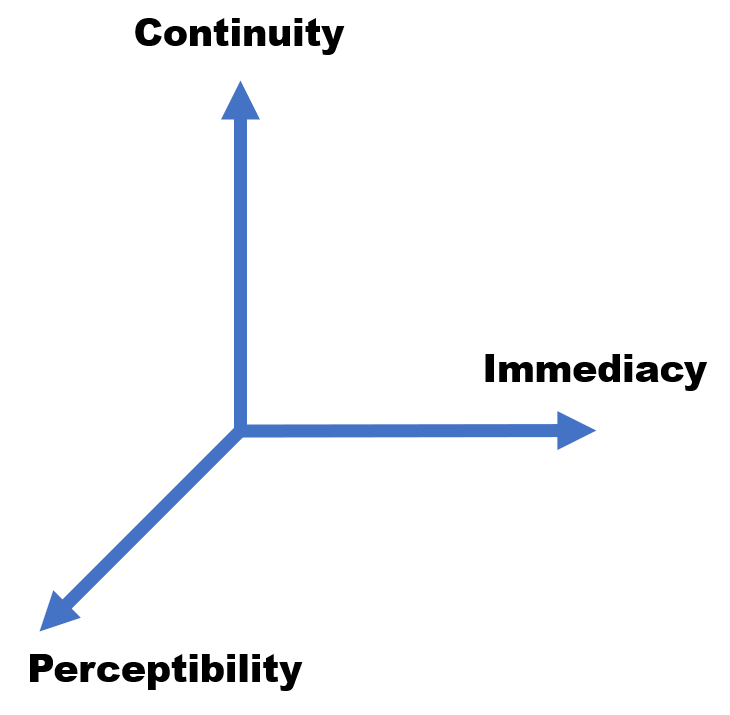}
    \\
    \caption{Three dimensions of liveness --- immediacy, continuity, and perceptibility}~\label{fig:liveness}
  \end{minipage}
\end{marginfigure}

First, we need to be able to measure the liveness of interactive systems to some extent to be able to compare interactive systems in terms of liveness. The three perspectives of liveness used in this paper can be one way to measure liveness in the interactive systems~\ref{fig:liveness}. The two temporal dimensions --- immediacy and continuity --- are straightforward to measure. The perceptibility --- the sense of being there --- can be tricky to quantify even thought it would be somewhat subjective based on the perception of spectators who experience events. In addition, how each dimension contributes to the overall liveness can be a challenging research work. Second, I wish to validate the benefits of having liveness in interactive systems for facilitating hybrid events and remote collaboration in general. The first goal is the prerequisite of the second goal. Once, liveness of interactive systems can be quantified, we will be able to assess the effects of having liveness conduct simple A/B testing by varying the level of liveness in the interactive systems that we use. 
I wish this paper can be a starting point to initiate the discussion of the technical definition of ``liveness'' in HCI.


\balance{} 

\bibliographystyle{SIGCHI-Reference-Format}
\bibliography{sample}

\end{document}